\documentstyle[12pt]{article}
\begin{document}
\newcommand{\bea}{\begin{eqnarray}}
\newcommand{\eea}{\end{eqnarray}}
\newcommand{\be}{\begin{equation}}
\newcommand{\ee}{\end{equation}}
\newcommand{\non}{\nonumber}
\newcommand{\ov}{\overline}
\global\parskip 6pt
\begin{titlepage}
\begin{center}
{\Large\bf Discrete Quantum Field Theories}\\
\vskip .25in
{\Large\bf and the Intersection Form}\\
\vskip .50in
Danny Birmingham \footnote{Supported by Stichting voor Fundamenteel
Onderzoek der Materie (FOM)\\
Email: Dannyb@phys.uva.nl}     \\
\vskip .10in
{\em Universiteit van Amsterdam, Instituut voor Theoretische Fysica,\\
Valckenierstraat 65, 1018 XE Amsterdam, \\
The Netherlands} \\
\vskip .50in
Mark Rakowski\footnote{Email: Rakowski@maths.tcd.ie}   \\
\vskip .10in
{\em School of Mathematics, Trinity College, Dublin 2, Ireland}  \\
\end{center}
\vskip .10in
\begin{abstract}
It is shown that the standard mod-$p$ valued intersection form
can be used to define Boltzmann weights of subdivision invariant
lattice models with gauge group $Z_{p}$. In particular, we discuss
a four dimensional model which is based upon the assignment
of field variables to the $2$-simplices of the simplicial complex.
The action is taken to be the intersection
form defined on the second cohomology group of the complex, with
coefficients in $Z_{p}$.
Subdivision invariance of the theory follows when the coupling constant
is quantized and the field configurations are restricted to those
satisfying a mod-$p$ flatness condition.
We present an explicit computation of the partition function for the
manifold $\pm CP^{2}$, demonstrating non-triviality.
\end{abstract}
\vskip .1in
\begin{center}
ITFA-94-19\\
June 1994
\end{center}
\end{titlepage}

\section{Introduction}
In \cite{BR1,BR2}, a general strategy for the construction of
subdivision invariant lattice models with $Z_{p}$ gauge group
has been presented. One of the main ingredients in these models
is the assignment of field variables (termed colours)
to simplices of various dimensions. Subdivision invariance is secured
by restricting to quantized values of the coupling parameter, and
summing over field configurations which satisfy a mod-$p$ flatness
condition, which are thus in correspondence with the cohomology classes
of the simplicial complex $K$.

In three dimensions, it was shown \cite{BR1} that the model of
Dijkgraaf-Witten \cite{DW} is defined in terms of a $1$-colour field $A$,
with an action of the form $A \cup \delta A$, where
the cup product and coboundary operator are defined in \cite{BR1}.
Of special interest is a model in four dimensions which relies
on the introduction of a $2$-colour $B$ and a $1$-colour $A$,
with an action of the form $B \cup \delta A$. Indeed, non-triviality
of this model has been established in \cite{BR2}.

In each of these examples, one notes a formal resemblance of the action
to the mod-$p$ intersection form. However, since these actions are
``kinetic''
in the sense that they contain the coboundary operator, this resemblance
is indeed only formal.

On the contrary, the purpose of the present note is to observe that one
can consider models defined with an action given in terms of the
standard mod-$p$ intersection form. The basic model which we shall
concentrate on here is defined in terms of a $2$-colour $B$
with an action of the form  $B \cup B$. In this case, subdivision
invariance requires the coupling to be parametrized as a $p$th root of
unity. Due to this fact, the action is identified with
the mod-$p$ intersection form on the second cohomology group of $K$
with coefficients in $Z_{p}$, evaluated on the fundmental homology
cycle of $K$.

\section{Subdivision Invariant Models and the Intersection Form}

For details of the general formalism we refer the reader to \cite{BR1,BR2}.
Let us begin by recalling the four dimensional model.
The Boltzmann weight of an ordered $4$-simplex $[0,1,2,3,4]$ is defined by:
\bea
W[[0,1,2,3,4]] &=& \exp\{\beta <B \,\cup \,
\delta A, [0,1,2,3,4]>\}  \non\\
&=&  \exp\{\beta\, B_{012}\, ( A_{23} + A_{34} - A_{24})\}
\;\;,     \label{bw1}
\eea
where $B$ and $A$ are $2$- and $1$-colour fields. When the gauge group
is taken to be $Z_{p}$,
a k-colour field is just an assignment of an integer in the set
$\{ 0, ... , p-1 \}$ to each of the k-simplices in the simplicial
complex.
Here, $\beta$ is the coupling parameter, and we shall also find
it convenient to use the scale $s = \exp[\beta]$.
One can now check the behaviour of this theory under subdivision
of the underlying simplicial complex.
It will suffice to recall here the behaviour when an additional
vertex $x$ is placed at the centre of the $4$-simplex
$[0,1,2,3,4]$, and linked to the other $5$ vertices.
The original $4$-simplex is then replaced by an assembly of five
$4$-simplices, written symbolically as:
\begin{eqnarray}
[0,1,2,3,4] &\rightarrow&
[x,1,2,3,4] - [x,0,2,3,4] + [x,0,1,3,4] \non\\
&-& [x,0,1,2,4] + [x,0,1,2,3]\;\;,
\end{eqnarray}
where we declare the new vertex $x$ to be the first in the
total ordering of all vertices.

One easily verifies that the Boltzmann weight behaves in the following
way under such a subdivision move:
\bea
& &W[[0,1,2,3,4]] s^{-<\delta B \,\cup\, \delta A, \, [x,0,1,2,3,4]>}
= W[[x,1,2,3,4]] \label{6W}   \\
& &W[[x,0,2,3,4]]^{-1}\, W[[x,0,1,3,4]]\,
W[[x,0,1,2,4]]^{-1}\, W[[x,0,1,2,3]]\;\;.\non
\eea
It is clear that subdivision invariance can be secured once
the insertion factor on the left hand side is trivialized. This
is accomplished by imposing a quantization of the coupling,
$s^{p^{2}} = 1$, and restricting the colourings
to those satisfying the ``flatness" conditions
\bea
[\delta B] = [\delta A] = 0\;\; .\label{flat1}
\eea
Thus, on the $2$-simplex $[0,1,2]$, the $1$-colour field is restricted by
\be
[\delta A]_{012} \equiv [A_{12} - A_{02} + A_{01}] = 0\;\;, \label{flat2}
\ee
while on the $3$-simplex $[0,1,2,3]$, the
restriction on the $2$-colour takes the form:
\be
[\delta B]_{0123} \equiv [B_{123} - B_{023} + B_{013} - B_{012}] = 0 \;\;.
\label{flat3}
\ee
Here the bracket notation indicates that the quantity is to be taken
modulo $p$ ($[x] = 0$ means $x = 0$  $mod-p$).
With these restrictions, the product $\delta B \cup \delta A$ is
a multiple of $p^{2}$ and hence the above insertion becomes unity.

It is also worth noting that the flatness conditions (\ref{flat1}) are
not the classical field equations which result from the action
$B\cup \delta A$. One is genuinely summing over field configurations
in the partition function
which can have non-zero field strength ($\delta A$ or $\delta B$).

The subdivision invariant Boltzmann weight for the $4$-simplex
$[0,1,2,3,4]$ is then given by:
\be
W[[0,1,2,3,4]] = \exp\{\frac{2 \pi i k}{p^{2}}\,
B_{012}\,( A_{23}\,+\, A_{34} \,-\,
[ A_{23} + A_{34}] )\}\;\;, \label{sdibw}
\ee
with $k \in \{0,1,\cdots,p-1\}$.

One other important feature of the Boltzmann weight is
that it possesses a local gauge invariance when the simplicial complex
is closed.
The gauge transformation of the A field defined on the ordered
$1$-simplex $[0,1]$ is defined by:
\be
A^{\prime}_{01} = [ A - \delta \omega ]_{01} =
[A_{01} - \omega_{1} + \omega_{0}]\;\;, \label{gt1}
\ee
where $\omega$ is a $0$-colour field defined on the vertices of the complex.
For the $2$-colour field $B$ defined on the
ordered $2$-simplex $[0,1,2]$, we have  a gauge transformation given by:
\be
B^{\prime}_{012} = [ B - \delta \lambda ]_{012} =
[B_{012} - \lambda_{12} + \lambda_{02} - \lambda_{01}]\;\;,
\label{gt2}
\ee
where $\lambda$ is a $1$-colour defined on $1$-simplices.

  Invariance of the theory under the above transformations is not
manifest, but requires both the quantization of the coupling parameter,
together with the restriction on the allowed field configurations. One
easily finds that under the transformation of $B$,
\bea
s^{B^{\prime} \cup \delta A} = s^{B\cup \delta A} \;
s^{- \delta \lambda \cup
\delta A} = s^{B\cup \delta A}\;s^{-\delta (\lambda \cup \delta A)} \;\; ,
\eea
where the first equality uses the fact that $\delta A$ is proportional
to $p$ due to the flatness constraint, and that $s$ is a $p^{2}$-root
of unity. Hence the Boltzmann weight is invariant up to a total boundary
term and the product of all these cancels for a closed oriented complex.
Invariance under the $A$ field transformation follows in
the same way if one first notes the simple identity,
\bea
s^{B\cup \delta A} = s^{- \delta B\cup A} \; s^{\delta ( B\cup A)} \;\; .
\eea

Similarly, in three dimensions, a Boltzmann weight of the form \cite{BR1}:
\bea
W[[0,1,2,3]] &=& \exp\{\beta <A \,\cup\, \delta A,[0,1,2,3]>\}\non\\
&=& \exp\{\beta \, A_{01}\, (A_{12} + A_{23} - A_{13})\} \;\;,
\label{dw}
\eea
leads to a subdivision invariant model, which is known as
the Dijkgraaf-Witten model \cite{DW}.

In each of these models one notices that the action is ``kinetic'' in the
sense
that the coboundary operator is involved. As a result, these actions
bear only a formal resemblance to the mod-$p$ valued intersection form.

Let us now deal with the main order of business which is to consider
models which are indeed based upon the intersection form.
In four dimensions, let us consider again the $2$-colour field $B$,
along with the Boltzmann weight
\bea
W[[0,1,2,3,4]] &=& \exp\{\beta <B \,\cup \,B, [0,1,2,3,4] >\}\non\\
&=& \exp\{\beta \, B_{012} \, B_{234}\} \;\;.
\label{bb}
\eea
The behaviour of this model under the subdivision move described above
is given by:
\bea
& &W[[0,1,2,3,4]] s^{-<(\delta B \,\cup\, B
+ B \,\cup\, \delta B),\, [x,0,1,2,3,4]>}
= W[[x,1,2,3,4]]   \\
& &W[[x,0,2,3,4]]^{-1}\, W[[x,0,1,3,4]]\,
W[[x,0,1,2,4]]^{-1}\, W[[x,0,1,2,3]]\;\;.\non
\eea
In this case, we see that if the coupling parameter is chosen to satisfy
$s^{p} = 1$, and the field configurations restricted to satisfy
the flatness conditions, we again have a subdivision invariant
Boltzmann weight.

The fact that the subdivision invariant Boltzmann weight requires
the scale parameter to a $p$th root of unity means it can equally well be
written as:
\be
W[[0,1,2,3,4]] = s^{[B_{012} \,B_{234}]}\;\;, \label{bbbw}
\ee
where the square brackets indicate that the product of B fields is now
taken modulo $p$. Thus, for a given
closed simplical complex $K$, the form of the action is specified
as the mod-$p$ intersection form on the second cohomology group (with
$Z_{p}$ coefficients) evaluated on the fundamental homology cycle of
$K$.

The verification of gauge
invariance, and independence of the choice of vertex ordering follows
as described in \cite{BR1,BR2}.

\section{The Partition Function for $CP^{2}$}
We continue in this section by evaluating the partition function
of the $B \cup B$ theory (equation (\ref{bb})) for complex projective space.
An economical simplicial complex for
the manifold $CP^{2}$ with  a minimal number of 9 vertices
has been given in \cite{tric}, and we label its vertices by elements
in the set $\{1,...,9\}$.  The complex is fully determined by
specifying the 4-simplices which are 36 in number and are given
explicitly by,
\bea
& &+ [1,2,4,5,6] + [4,5,7,8,9] + [1,2,3,7,8] + [2,3,4,5,6] \non\\
& &+ [5,6,7,8,9] + [1,2,3,8,9] - [1,3,4,5,6] - [4,6,7,8,9] \non\\
& &- [1,2,3,7,9] - [1,2,4,5,9] - [3,4,5,7,8] - [1,2,6,7,8] \non\\
& &- [2,3,5,6,7] - [1,5,6,8,9] - [2,3,4,8,9] - [1,3,4,6,8] \non\\
& &- [2,4,6,7,9] - [1,3,5,7,9] + [2,3,4,6,9] + [3,5,6,7,9] \non\\
& &+ [1,3,6,8,9] + [1,3,4,5,7] + [1,4,6,7,8] + [1,2,4,7,9] \non\\
& &- [1,2,5,6,8] - [2,4,5,8,9] - [2,3,5,7,8] + [1,3,5,6,9] \non\\
& &+ [3,4,6,8,9] + [2,3,6,7,9] + [1,2,4,6,7] + [1,4,5,7,9] \non\\
& &+ [1,3,4,7,8] - [2,3,4,5,8] - [2,5,6,7,8] - [1,2,5,8,9] \;\;,
\label{cp2}
\eea
where the signs denote the relative orientations of each simplex.
The corresponding simplicial complex for the orientation reversed
complex projective space, $-CP^{2}$, is obtained by reversing the signs
of each $4$-simplex in (\ref{cp2}).

The number of simplices of each dimension contained in $K$ is:
\bea
& &0-simplices: 9 \non\\
& &1-simplices: 36 \non\\
& &2-simplices: 84 \non\\
& &3-simplices: 90 \non\\
& &4-simplices: 36 \;\;.
\eea

We must first solve the flatness conditions (\ref{flat3}) subject
to the gauge equivalence (\ref{gt2}). Here, we have $84$  $B$ fields
subject to $90$ constraints. As always when solving these constraints,
it is convenient to make use of the gauge freedom,
and one readily determines that a maximal number of $28$
fields can be gauge fixed. The resulting solution is expressed
in terms of a single independent field variable
$x \in \{0,1,\cdots , p-1\}$.

As described in \cite{BR2}, a certain scaling factor is required
in the definition of the subdivision invariant partition function.
Since in the present case we have only a single $2$-colour field,
the partition function is defined by:
\be
Z = \frac{1}{|G|^{N_{1} - N_{0}}} \sum_{flat} W[K]\;\;.
\ee

For the gauge group $Z_{2}$, we have two independent solutions
and the partition function evaluated at the
non-trivial root of unity, $s = -1$ in equation (\ref{bbbw}),
takes the form:
\be
Z[\pm CP^{2}, Z_{2}] =  \frac{1}{2^{27}}\, 2^{28}\, (1 - 1) = 0\;\;,
\ee
where the factor $2^{28}$ accounts for the gauge equivalent copies
of the solution.

With gauge group $Z_{3}$, there are three gauge inequivalent
solutions to the flatness constraints and the partition function
evaluated at $s = \exp[2\pi i/3]$ is given by:
\bea
Z[+CP^{2}, Z_{3}] &=& \frac{1}{3^{27}} \, 3^{28} \, (1
+ 2 \, \exp[-\frac{2 \pi i}{3}]) = - 3 \, \sqrt{3} \, i \;\;,\non\\
Z[-CP^{2}, Z_{3}] &=& \frac{1}{3^{27}} \, 3^{28} \, (1 +
2 \, \exp[\frac{2 \pi i}{3}]) = + 3 \, \sqrt{3} \, i \;\;.
\eea
We see therefore that the partition function is sensitive to
orientation. Of course, a reversal of orientation always takes
$Z$ into its complex conjugate.

For the sake of comparison, and in order to establish the normalization,
it is useful to compare the above result with that of the 4-sphere $S^{4}$.
A simplicial complex is given simply by the boundary of a single 5-simplex;
here one has $N_{0}=6$, $N_{1}=15$, $N_{2}=20$, $N_{3}=15$, and $N_{4}=6$.
The partition function is,
\bea
Z[S^{4},G] = \frac{1}{|G|^{15-6}} \, |G|^{10} \, 1 = |G| \;\; .
\eea

\section{Observables}

In any quantum field theory, one typically defines a set of gauge invariant
observable operators which yield additional interesting information
beyond the partition function.
To illustrate the general structure of these observables, let us consider
the $B \cup \delta A$ model in four dimensions.
It is evident that one can construct a gauge invariant observable
involving the $A$ field as follows:
\be
W(\gamma_{1}) = \exp\{ \beta^{\prime} < A, \gamma_{1} >\}\;\;,
\ee
where $\gamma_{1}$ symbolically denotes a homology $1$-cycle
in the complex $K$. Here, $\beta^{\prime}$ is a coupling parameter,
and we shall also use the notation $s^{\prime} = \exp[\beta^{\prime}]$.
In order for this quantity
to be gauge invariant, and dependent only on the homology class of
$\gamma_{1}$,
we are again focred to quantize this coupling. From
(\ref{gt1}), we see that when $(s^{\prime})^{p} = 1$,
the gauge transformed observable is given by
\bea
W^{\prime}(\gamma_{1}) &=& \exp\{\beta^{\prime}
<[A - \delta \omega]\, , \gamma_{1}>\}\non\\
&=& W(\gamma_{1}) (s^{\prime})^{ - <\delta \omega \, , \gamma_{1}>} \non\\
&=& W(\gamma_{1})\;\;.
\eea
This is a simple consequence of the fact that $([x + y] - x - y)$ takes
the value $0$ or $p$, and that $\gamma_{1}$ is a homology cycle.

A similar arugument shows that $W(\gamma_{1})$ depends only on the
homology class of the cycle $\gamma_{1}$. Consequently, such an observable
is trivial when $\gamma_{1}$ is homologically trivial.

Additionally, a gauge invariant observable involving the $B$ field
can be defined by
\be
W(\gamma_{2}) = \exp\{\beta^{\prime} <B\, , \gamma_{2} >\}\;\;,
\ee
where one verifies that the observable depends only on the
homology class of the $2$-cycle $\gamma_{2}$.
One can also contemplate the evaluation of
an observable such as
\be
W(\gamma_{3}) = \exp\{\beta^{\prime} < B\,\cup\,A\, , \gamma_{3} > \}\;\;,
\ee
where $\gamma_{3}$ is a homology $3$-cycle.

Finally, we should mention that interesting structures may also emerge by
considering the coupling of the intersection and kinetic models.
In this respect, one may view the addition of the $B \cup B$ term
to the kinetic $B \cup \delta A$ action as the evaluation of
the observable $B \cup B$ in the kinetic theory.

\section{Remarks}
Our construction gives a topological quantum field theory (TQFT) flavor
to the intersection form, a well known topological invariant. One can
expect that all
the axioms \cite{At} of a TQFT will be obeyed. Clearly there is nothing
sacred about the particular example we considered in four dimensions,
and the same construction can be undertaken in any dimension, possibly
with a mix of fields of various colour types. Subdivision invariance
is achieved through a quantization of the coupling parameter, together
with a restriction of the field configurations in the theory. One might
also view this $B\cup B$ theory as an additional coupling (or observable)
to the ``kinetic'' term $B\cup \delta A$ introduced in \cite{BR1,BR2}.

\end{document}